\documentclass[twocolumn,showpacs,superscriptaddress,groupedaddress,floatfix]{revtex4-1}

\usepackage{color}
\usepackage[]{graphicx}  
\usepackage{bm}        
\usepackage{amssymb}
\usepackage{gensymb}
\usepackage{amsmath}   
\usepackage{float}
\usepackage[english]{babel}
\usepackage{color}
\setcitestyle{super}

\addto\captionsenglish{}

\def\perkg{\mathrm{kg^{-1}}}
\def\permm{\mathrm{m^{-2}}}
\def\permmm{\mathrm{m^{-3}}}
\def\persec{\mathrm{s^{-1}}}

\def\2D{\mathrm{2D}}
\def\3D{\mathrm{3D}}

\def\cpunits{\mathrm{J~kg^{-1}~K^{-1}}}
\def\perhour{\mathrm{hour^{-1}}}

\def\N2{$\mathrm{N}_2$}

\def\cmr{(  q/m )}

\def\h2o{$\mathrm{H}_2\mathrm{O}$}
\def\co2{$\mathrm{CO}_2$}
\def\o2{$\mathrm{O}_2$}

\def\psat{p_{\mathrm{sat.}}}
\def\Qabs{Q_{\mathrm{abs.}}}
\def\Abse{| e |}
\def\Tm{T_{\mathrm{m}}}

\begin{document}

\widetext
\title{Observation of undercooling in a levitated nanoscale liquid Au droplet}
\author{Joyce Coppock}
\email{jec@umd.edu}
\author{Quinn Waxter}
\author{Robert Wolle}
\affiliation{University of Maryland, College Park, MD, 20742, USA}
\affiliation{Laboratory for Physical Sciences, 8050 Greenmead Dr., College Park, MD, 20740, USA}
\author{B. E. Kane}
\affiliation{Joint Quantum Institute, University of Maryland, College Park, MD, 20742, USA}
\affiliation{Laboratory for Physical Sciences, 8050 Greenmead Dr., College Park, MD, 20740, USA}
\date{\today}

\begin{abstract}
\noindent
We investigate melting and undercooling in nanoscale (radius $\sim$100 nm) gold particles that are levitated in a quadrupole ion (Paul) trap in a high vacuum environment. The particle is heated \textit{via} laser illumination and probed using two main methods. Firstly, measurements of its mass are used to determine the evaporation rate during illumination and infer the temperature of the particle. Secondly, direct optical measurements show that the light scattered from the particle is significantly different in its liquid and solid phases. The particle is repeatedly heated across its melting transition, and the dependence of heating behavior on particle size is investigated. Undercooling -- the persistence of a liquid state below the melting temperature -- is induced \textit{via} multi-stage laser pulses. The extent of undercooling is explored and compared to theoretical predictions.  
\end{abstract}

\maketitle

Undercooling is the phenomenon where a liquid can be cooled below its melting temperature and is frequently limited by nucleation of the solid phase around impurities or solid materials in contact with the melted material. Levitation (or more broadly, containerless processing) has been used for over fifty years to measure $10^{-6}-10^{-3}$ kg scale samples in a regime where interactions with supporting materials are completely eliminated.\cite{Herlach1991,Herlach2015} These techniques have been successfully applied to probe a wide variety of materials, including refractory metals in an ultra high vacuum (UHV) environment.\cite{Paradis2005,Klein2009}

In many cases nucleation that limits undercooling is thought to originate in the bulk of the material.\cite{Bokeloh2011,Bokeloh2014} Consequently, small samples can lead to the reduction of nucleation sites, or even their elimination if the site is a trace impurity. Early measurements of undercooling were undertaken using an optical microscope with sample sizes as small as 30 $\mu$m.\cite{Turnbull1950} More recently, chip based microcalorimetry has been used to probe undercooling in samples as small as 10 $\mu$m,\cite{Yang2011,Gao2009} with mass 
$m<10^{-11}~$kg. Microscopic samples probed with these techniques must be attached to a substrate, however.

In order to extend measurements of levitated materials to the nanoscale, we have developed an ion trap apparatus capable of measurements in high vacuum for extended periods.\cite{Nagornykh2015,Coppock2017} We have used this system to study $2r=200$ nm diameter Au nanospheres ($m\equiv m_0=8\times10^{-17}~$kg) and shown that measurement of mass loss can be used for accurate determination of the temperature $T$ of the levitated particle near its melting point $\Tm$=1337 K.\cite{Coppock2021} By controlling the power of a laser illuminating the sample, we can heat the particle above $\Tm$ and subsequently cool it to an undercooled state. Determination of the phase of the particle is made from both differences in the absorption (inferred from temperature differences) and scattering of the laser light from the particle. We have observed undercooling to a maximum of 212 K below $\Tm$, a value very similar to that reported in samples with $m=10^{-4}~$kg.\cite{Wilde2006} Naive scaling of the nucleation rates obtained in macroscopic samples \cite{Bokeloh2014} would predict that we should observe undercooling to $\Tm-T\simeq$270 K. We will discuss possible explanations for this discrepancy after we have presented our data.

In the trap, the particle is illuminated by a 532 nm laser with power controlled by a Pockels cell with a $<10^{-3}~$s response time. The power of the beam exiting the vacuum chamber is measured by a fast photodiode. We collect data using an optical power density $S$ ranging from $10^3-10^5$~W$\permm$.

Most of the measurements we describe below are inferred from the charge to mass ratio $q/m$ of the trapped particle, determined by measurements of the frequencies of oscillation of the particle in the trap. To derive $q/m$ from the trap frequencies, formulas described by Illemann\cite{Illeman2000} were used. In our measurements of $q/m$, typical precision after averaging for about 1 minute is $~10^{-4}$ C $\perkg$. Our $q/m$ values were calibrated previously\cite{Coppock2018} using time of flight measurements, and are accurate to $\sim 1\%$.

Typical trapped particles have a total charge number of $q/|e|$=1000-1500. Our experiments rely on the fact that, after a heat treatment procedure (described in the Experimental Methods section), the charge number is mostly stable.  The particle occasionally ($\sim0.1-1 ~\perhour$) loses single elementary charges that allow $q$ and $m$ separately to be determined from the observed change $\delta \cmr$ in $q/m$:

\begin{figure}
\begin{center}
\includegraphics[scale=1,draft=false]{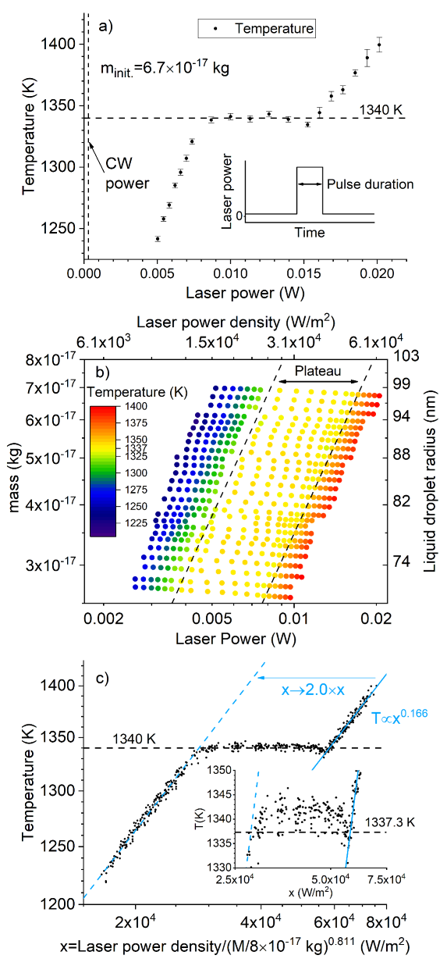} 
\end{center}
\caption{
(a) $T$ behavior of a particle derived from a single pulse power sweep.  The particle is exposed to single heating pulses from the laser with duration varying from 2.23 s (lowest power) to 0.07 s (highest power).  (b) Color-mapped plot of $T$ on logarithmic axes of laser power and particle mass. (c) Collapsed data plotted on log axes.  Inset shows magnified data near the plateau.
}
\label{Fig1}
\end{figure}

\begin{figure}
\begin{center}
\includegraphics[scale=1,draft=false]{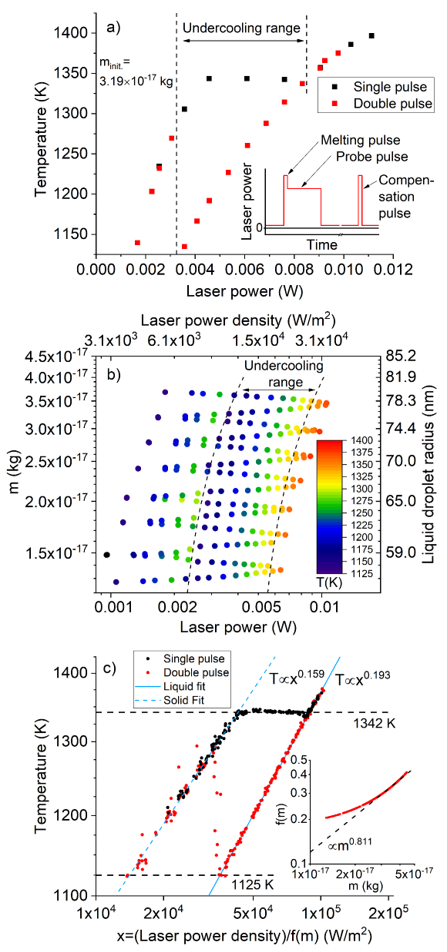} 
\end{center}
\caption{
(a) $T$ behavior deduced from measurements of single and double pulses, demonstrating undercooling.  (b) Colormapped $T$ data for double pulses only, plotted on log axes of laser power and $m$.  (c) Collapsed data  on log axes for single and double pulses.  Inset shows the scaling function used.
}
\label{Fig2}
\end{figure}

\begin{equation} \label{eqn:a}
q = \frac{\Abse \cmr}{\delta \cmr}
\end{equation}
and
\begin{equation} \label{eqn:b}
m = \frac{\Abse}{\delta \cmr}.
\end{equation}
Also, subsequent to thermal treatment, discharge is $not$ a thermally activated process, and consequently short ($\sim$1 s)  heat pulses applied to the particle cause a change in $q/m$ usually entirely attributable to mass gain or loss:
\begin{equation} \label{eqn:c}
\frac{\delta m}{m}=-\frac{\delta \cmr}{\cmr}.
\end{equation}
In the instances when a discharge event coincides with a heat pulse, it is usually possible to compensate for it if the expected $\delta m$ is known approximately, since $\delta q$ must come in multiples of $|e|$.

At sufficiently high $T$, particles in our trap lose mass by Au evaporation:\cite{Langmuir1913}
\begin{equation} \label{eqn:d}
\frac{\dot{m}}{\mathrm{area}}=  -\psat \sqrt \frac{m_{\mathrm{Au}}}{2 \pi k_B T}.
\end{equation}
Here, $\dot{m}=dm/dt$, $\psat$ is the vapor pressure, and 
$m_{\mathrm{Au}}$=3.271$\times 10^{-25}$ kg is the mass of an Au atom. The validity of Eq. \ref{eqn:d} requires that the sticking coefficient of Au vapor impinging on the surface is unity.

If the shape of the particle is spherical and it is assumed to have constant density $\rho$, then:

\begin{equation} \label{eqn:e}
\frac{\dot{m}}{m^{2/3}} =-\psat \times \sqrt[3]{{\frac{36 \pi }{\rho^2}}}\times  \sqrt \frac{m_{\mathrm{Au}}}{2 \pi k_B T}.
\end{equation}
The $T$ dependence of $\psat$,\cite{Alcock1984} as well as the solid and liquid densities $\rho_{\mathrm{s}}$\cite{Pamato2018} and $\rho_{\mathrm{l}}$\cite{Paradis2008}, are all known to high accuracy.  Thus, knowledge of the parameters on the left hand side of Eq. \ref{eqn:e} enables the determination of $T$.

In our experiments the continuous wave (CW) illumination power of the laser, $\cong$ 300 $\mu$W, is chosen to provide high measurement sensitivity while causing negligible mass evaporation. Data is collected by pulsing the laser to high power for durations ranging from 0.07 s to 30 s. The duration depends on the pulse power and is chosen so that $\delta m/m<10^{-3}$. As will be demonstrated below (in the subsection titled "Determination of refreezing time"), the pulse durations are all much longer than the time taken for the particle to thermalize, so it is valid to assume that the particle $T$ is constant during the pulse. To take data, the laser is pulsed in 10-20 minute intervals to allow for determination of $q/m$ by averaging between pulses. Occasional discharge events are also recorded and used to track the mass and charge of the particle during the course of the experiment.

\section*{Results and Discussion}
\subsection{Heating and melting using single pulses}

Data for a set of single laser pulses with powers appropriate to heat the particle up to the vicinity of the Au melting temperature $\Tm$=1337 K is shown in Fig. \ref{Fig1}a. The error bars reflect the uncertainty of the mass measurements made before and after each pulse. As has been observed previously,\cite{Coppock2021} a plateau appears in the data very near to $\Tm$, likely a consequence of the reduction in the absorption efficiency $\Qabs$ when solid Au melts, and the surface plasmon resonance, which has a large effect on $\Qabs$ at $\lambda$=532 nm, is suppressed.\cite{Gerasimov2016,Ershov2017}  For illumination powers on the plateau, the particle is in a partially melted state.

Because each laser pulse vaporizes $\sim 0.1\%$ of the mass of the particle, taking a thorough data set requires substantial loss of mass. Fig. \ref{Fig1}b shows 489 color-mapped $T$ data data points as functions of the laser power and particle mass, plotted on a logarithmic scale. As the particle becomes smaller, the laser power necessary to reach the plateau declines, but the plateau $T$ remains very close to the Au $\Tm$ bulk value. Over the limited range of $T$ and $m$ that is experimentally probed, the data can be effectively collapsed onto a scaled x-axis with
$x=S/[m/m_0]^{0.811}$. Isotherms obtained from this equation are plotted in \ref{Fig1}b and the entire scaled data set is plotted in Fig. \ref{Fig1}c. Below the plateau, the $T$ dependence of all data is well fit by $T^{1/6}$. The plateau spans about a factor of two in pulse power.  Interestingly, there is a slight decrease of the plateau $T$ at powers just below those required to fully liquefy the particle.

It is remarkable that Eqs. \ref{eqn:d} and \ref{eqn:e} predict $T$ as accurately as we have observed. The 2 K difference between our measured value of $T$ and the value of $\Tm$=1337.3 K corresponds to a $5\%$ error in the evaporation rate, which could arise from errors in our determination of $q$ and $q/m$. It should be noted, however, that Eq. \ref{eqn:e} is derived assuming a spherical particle, and the formula will overestimate $T$ of non-spherical particles, due to the increased evaporation off of their necessarily larger surface area. Thus, it is 
possible the shape of the mostly solid particles contributes to the small deviation of the plateau $T$ from the bulk value.

\subsection*{Observation of undercooling using double pulses}

In order to observe undercooling, it is necessary first to bring the particle fully to its liquid state and then cool it into the regime where $T<\Tm$. We accomplish this with a double pulse: the first (melting) pulse typically heats the particle to $\sim1400$ K for 0.07 seconds. The second (probe) pulse has varying power and duration appropriate for reaching and measuring its expected $T$ during the pulse (see Fig. \ref{Fig2}a inset). As was the case for single pulses, $T$ is determined from the mass loss rate during the probe pulse. Because of the long averaging times necessary to accurately measure $q/m$, it is only possible to measure $\delta \cmr$ for a double pulse, which will include a substantial contribution from the initial melting pulse. To remove this contribution we also measure $\delta \cmr$ for a compensation pulse with the same parameters as the melting pulse and then subtract the measured $\delta \cmr$ for the value obtained from the double pulse. The measuring procedure thus requires a sequence of double pulses, compensation pulses, and\textemdash for concomitant measurement of the non-undercooled particle\textemdash occasional single pulses, each separated by $\sim$15 minute averaging periods.

Data for a single sweep of the laser power is shown in Fig. \ref{Fig2}a. The double pulse data follows the liquid state trend line down to $T<1150$K, a point where the single pulse temperatures are below the plateau. A full sequence of power sweeps is shown in 
Fig. \ref{Fig2}b, with only the double pulse data plotted. As was the case with the single pulse data, the power required to achieve a given $T$ declines as the particle becomes smaller. Data points are more sparse for double pulses than they were for single pulses (Fig.\ref{Fig1}b ) because of the extra compensation pulses and single pulses needed for the data set. Collapsing the data onto a single two dimensional plot is possible, but the scaling function, $f(m)$, is more complicated\cite{Fit} than the power law function used in the single pulse data. $f(m)$ is plotted in the inset to Fig.\ref{Fig2}c, and its deviation from a power law only occurs when $m$ is lower than the values presented in Fig. \ref{Fig1}.

The scaled and collapsed data for both single and double pulses is shown in Fig. \ref{Fig2}c. The liquid state trend line for the double pulse data persists down to $T$=1125 K, or 212 K below $\Tm$. For undercooled data points, parameters used to determine the right hand side of Eq. \ref{eqn:e} were values extrapolated below $\Tm$ using formulas for $\psat$\cite{Alcock1984} and $\rho$\cite{Paradis2008} obtained from data taken when $T>\Tm$. The use of solid state parameters would lead to a $\sim$10 K increase in the minimum undercooling $T$.

The undercooling data provides information on the liquid state over a greater temperature range than the single-pulse data in Fig. \ref{Fig1}. In the liquid regime the $T$ dependence on laser pulse power is well fit by a power law $T=(C_0 S)^{1/\alpha}$ with $\alpha$=5.18, somewhat different from the exponent ($\cong 1/6$) that best fits the solid data. We attribute this difference to the absence of the surface plasmon resonance and smaller electrical conductivity of the liquid compared to the solid at the same $T$.

\subsection*{Determination of refreezing time}
\label{sec:time}

\begin{figure}
\begin{center}
 \includegraphics[scale=1,draft=false]{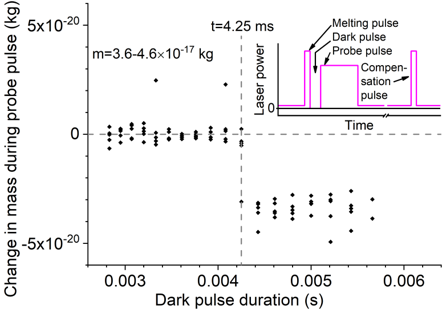}
\end{center}
  \caption{A three pulse sequence is used to determine the time for a liquid particle to refreeze. A dark pulse of variable duration is inserted between a melting pulse and a probe pulse, and a compensation pulse with the same parameters as the melting pulse is measured separately. Inferred mass loss during the probe pulse shows a sharp transition when the dark pulse duration exceeds 4.25 ms.}
\label{Fig3}
\end{figure}

To directly probe the time necessary to refreeze a particle in a liquid state after the laser power is switched off, a three-pulse power sequence was used (Fig. \ref{Fig3} inset). A dark pulse of variable duration was inserted between the initial melting pulse with .07 s duration and a probe pulse with a duration of 0.25 s. The melting pulse typically heated the particle to 1400 K, while the probe pulse power was chosen so as to heat the particle to the low power edge of the plateau.  At this power the solid particle $T$=1340 K, while the liquid $T\approx$1175 K (Fig. \ref{Fig2}c). At these respective temperatures $\psat$ is 100$\times$ greater for the solid than for the liquid, and the state of the particle during the probe pulse can be easily deduced from measurements of the mass loss during the pulse sequence if the effect of the melting pulse is accounted for with a separate compensation pulse.

Data (Fig. \ref{Fig3}) shows that the particle becomes solid if the duration of the dark pulse exceeds 4.25 ms. By assuming the absorbed power is equal to $\Qabs(\lambda$=532$~\mathrm{nm}) \times \pi r^2 S$ and that the emitted and absorbed power must be equal at constant $T$, it is possible to derive a time dependent cooling curve by integrating the function fitting the liquid data in Fig. \ref{Fig2}c. The time $t$ to cool from an initial temperature $T_0$ to a final temperature $T_f$ is given by:

\begin{equation} \label{eqn:g}
\Qabs t=\frac{c_p C_0}{(\alpha-1)f(m)}\left ( \frac{16 \rho^2 m}{9 \pi} \right )^{1/3} \left\{T_f^{1-\alpha}-T_0^{1-\alpha} \right\},
\end{equation}
where $c_p \cong 165~\cpunits$ is the specific heat \cite{Wilde1996,Khvan2020} of liquid Au in the regime of undercooling. $C_0$(=1.97$\times10^{11}$), $\alpha$, and $f(m)$\cite{Fit} are all obtained from the liquid data fit in Fig. \ref{Fig2}c. We have neglected heating from the surrounding environment (at 300 K) in the derivation of Eq. \ref{eqn:g}. For $T_0$=1400 K, $T_f$=1125 K, $m=4\times10^{-17}$ kg, and $t$=4.25 ms, we obtain $\Qabs(\lambda$=532$~\mathrm{nm})$=0.90, a value  in reasonable agreement with values ($\simeq$0.75) reported for somewhat smaller particles.\cite{Gerasimov2016,Kallel2017}

\subsection*{Direct optical determination of Au particle phase}
\label{sec:scatter}

\begin{figure}
\begin{center}
 \includegraphics[scale=1,draft=false]{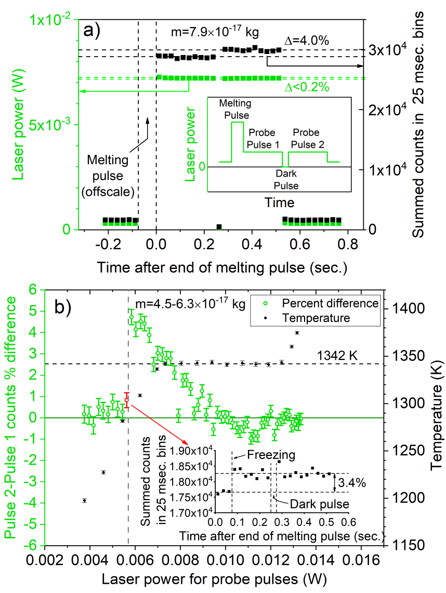}
\end{center}
  \caption{(a) Inset: Pulse sequence used for optical scattering measurements at 532 nm. After the particle is melted, scattering is measured during two probe pulses, separated by a dark pulse with sufficient duration to freeze the particle if it is in a liquid state.  Main figure shows data where there is a 4\% difference in scattering between the final solid state and the initial liquid state. (b) Dependence of the scattering difference signal on probe pulse power. Inset shows a data point where spontaneous freezing occurred during the first probe pulse.}
\label{Fig4}
\end{figure}

A significant disadvantage to using mass determinations for detecting undercooling is that the measurement process is slow, typically only a few data points per hour. Data taking would be much faster if there were a direct optical means for determining whether the particle is solid or liquid. Our current experimental setup is capable of detecting approximately 0.1\% of the 532 nm photons scattered by the particle, assuming isotropic scattering and accounting for the quantum efficiency of the detectors. At the powers where the particle is undercooled, we observe
a $\sim10^6$ counts $\persec$ scattered light signal. Thus, we should be capable of seeing very small changes in the scattered light signal if solid and liquid scattering differ.

The measurement protocol used to look for differences in scattered light is shown in the inset to Fig. \ref{Fig4}a. A single melting pulse is followed by two probe pulses, separated from each other by a dark pulse of sufficient duration (25 ms) to freeze the particle if it is in a liquid state. Counter data summed in 25 ms bins from a single pulse sequence is shown in the main figure, where the second (solid) pulse  produces a 4\% larger scattered signal than the first (liquid) pulse. We simultaneously measured the powers of the pulses, and they differ by $<0.2\%$.

The counter signal difference between the second and first probe pulse is plotted in Fig. \ref{Fig4}b as a function of probe pulse power, along with $T$ reached at this power using  single pulses. The difference signal is strongest at low powers at the onset of undercooling and diminishes essentially to zero at the high powers near where the particle is completely liquid. The data is consistent with a picture in which the particle is composed of a liquid outer shell surrounding a solid core,\cite{Guenther2014,Font2013,Coppock2021} and as the liquid layer surrounding the solid increases in thickness, the scattering difference between a partially solidified particle and an undercooled liquid droplet goes to zero.

Data in the inset to Fig. \ref{Fig4}b shows the direct observation of particle freezing during the first probe pulse. Mass loss during the probe pulses is dominated by the time the particle is in the solid (high $T$) phase. Determination of the particle phase in $\sim$25 ms means that mass loss per freezing event could be minimized in future experiments by interrupting the beam rapidly after the particle has solidified. Using such a protocol thousands of measurements of the freezing transition time could be made rapidly before mass loss becomes significant compared to the mass of the particle.

\subsection*{Size Effects}

In the interpretation of our data, we have neglected size-dependent properties of the particle that will become increasingly important if particles are smaller than the $r$=50-100 nm scale that we have investigated. Firstly, the finite curvature of the surfaces of small droplets produces an internal pressure $p_i$ inside the drop:

\begin{equation} \label{eqn:h}
p_{i}=\frac{2\gamma_l}{r},
\end{equation}
where $\gamma_l$ is the liquid surface energy.  
$p_{i}$ leads to an increase in $\psat$\cite{Moore1962}

\begin{equation} \label{eqn:i}
\frac{\delta \psat (r)}{\psat} = \frac{2 \gamma m_{\mathrm{Au}}}{r \rho k_B T},
\end{equation}
where we have assumed $\delta \psat (r)/\psat \ll $1. For liquid droplets near $\Tm$:
$\gamma=\gamma_l=1.135$ J $\permm$, \cite{Guenther2014} and 
$\rho=\rho_l$=17,400 kg $\permmm$.\cite{Paradis2008} For $r$=50 nm, 
$\delta \psat (r)/\psat$=0.04. Inclusion of this term in our determination of $T$ would reduce the result by $\cong$1 K and consequently has been neglected.

The second size effect that we have ignored is the change in $\Tm$ for nanoscale particles:\cite{Buffat1976,Guenther2014}
\begin{equation} \label{eqn:j}
\frac{\delta \Tm (r)}{\Tm} =  \frac{2}{r \rho_s H_m}
\left\{\gamma_l\left(\frac{\rho_s}{\rho_l}\right)^{2/3}-\gamma_s\right\},
\end{equation}
where $H_m$ the Au enthalpy of melting =$6.3\times10^4$ J $\perkg$ and 
$\rho_s$=18,300 kg $\permmm$. Eq. \ref{eqn:j} closely fits prior experimental data\cite{Guenther2014}  when the solid surface energy $\gamma_s$=1.4 J $\permm$ and predicts 
$\delta \Tm$(50 nm)$=-13$ K. We have not yet observed a size dependence of the melting plateau in our data.

\section*{Conclusions}
Perhaps the most surprising feature of our data is that the maximum undercooling we observe is essentially the same as that measured in samples with almost 10 orders of magnitude larger $m$. One explanation is that there are impurities present in our samples that are the nucleation centers for solidification. Impurities are the likely source of variance in achievable undercooling in experiments on macroscopic samples.\cite{Bokeloh2014,Wilde2006} The materials used in those studies are $99.9999\%$ pure, whereas our nanoparticles are manufactured by using citric acid to reduce gold chloride salt of $99.9\%$ purity.\cite{Lunanano} It is also possible that the \o2 necessary for our data taking may be a factor limiting undercooling.\cite{Wilde2006}

Because the particles we study are confined in an ion trap, the particles must inevitably be charged. Our data was taken with $q/|e|$ ranging from 500-1500, and we have observed no effect of particle charge on any of our measurements. We cannot rule out that surface charge could play a role in solid nucleation, however.

We note that for droplets with $r$=100 nm, $p_{i}$ determined from
Eq. \ref{eqn:h} is 23 MPa. Pressure has been proposed to promote the growth of solid nuclei,\cite{Silva2021,Kalyanaraman2008} and internal pressures typical in our experiments may significantly increase nucleation rates.\cite{Han2012} This effect could consequently provide an explanation for why undercooling is not greater in our samples.

Future experiments could greatly improve our understanding of undercooling in the nanoscale regime. Fast optical measurements to determine particle phase, highlighted in Fig. \ref{Fig4}, are most effective near maximum undercooling, and will facilitate probes of kinetics in this regime.  Measurements of samples with much better purity should be possible. Lastly, our techniques can be extended to a wide variety of materials in which the mass loss from evaporation in the regime of undercooling is neither too large nor too small.

\section*{Experimental Methods}
The ion trap apparatus and the techniques used to introduce Au nanoparticles into the trap for analysis have been described in previous publications.\cite{Nagornykh2015,Nagornykh2017,Coppock2017,Coppock2021} Purchased Au nanospheres\cite{Lunanano} with nominal diameter $2r$=200 nm are thoroughly rinsed and injected into an ion trap using an electrospray ion source. After initial characterization, the particles are transferred to a second trap, where measurements in high vacuum ($\geq10^{-8}~$Torr) can be performed for extended periods (weeks).

After transfer, particles typically have $q/m \simeq$5 C $\perkg$. To prepare the particles for measurements, a thermal treatment step is necessary, in which the laser power is gradually increased to a power density around 2000 W/m$^2$.\cite{Coppock2021} This removes contamination from the particle surface and leads to an ultimate $q/m$ of 2-3 C $\perkg$, which corresponds to a charge number $q/|e|$=1000-1500 for Au particles of nominal 100 nm radius and mass $m_0=8\times10^{-17}$ kg.  

In order to maintain stability of the trapped particle for extended periods, the particle center of mass motion in all three dimensions is cooled using the parametric feedback technique,\cite{Gieseler2012,Nagornykh2015} although cooling is modest (residual thermal motion of the particle is on the order of $\mathrm{T}=10-100$ K) owing to the necessary low laser powers required for our experiments. Additionally, DC electric fields on the particle are minimized using active feedback.\cite{Eltony2013,Nadlinger2021}

We have been able to obtain good data only when the particle is in the presence of \o2. We have speculated\cite{Coppock2021} that \o2 removes C that accumulates on the particle surface from CO and \co2 contaminants in the vacuum chamber. The presence of C on the surface reduces the sticking coefficient and Au evaporation rate at a given $T$ and invalidates Eqs. \ref{eqn:d} and \ref{eqn:e}. Consequently all data presented above was taken in \o2 with $p=2-3\times 10^{-6}$ Torr.

\section*{Acknowledgements}

This work was supported by the Laboratory for Physical Sciences, Contract \#H9823017C0194.

\bibliography{undercool}

\begin{mcitethebibliography}{37}
\bibitem[Fit()]{Fit}
$f(m)=(m/m_0)^\beta\times \exp\{\mu(-1+m/m_0)\}$, where $\beta$ and $\mu$ are
  fit parameters. In Fig. \ref{Fig2}c, $\beta$=-0.114 and $\mu$=2.137.\relax
\mciteBstWouldAddEndPunctfalse
\mciteSetBstMidEndSepPunct{\mcitedefaultmidpunct}
{}{\mcitedefaultseppunct}\relax
\EndOfBibitem
\end{mcitethebibliography}

\end{document}